\documentclass[useAMS,usenatbib]{mn2e}
\usepackage{times}
\usepackage{amssymb}
\usepackage{epsfig}
\usepackage{lscape,graphicx}
\title[Galaxy Environments in the UKIDSS Ultra Deep Survey (UDS)]{Galaxy Environments in the UKIDSS Ultra Deep Survey (UDS)}
\author[R. W. ~Chuter et al.]{R. W. Chuter$^{1}$\thanks{E-mail:ppxrwc@nottingham.ac.uk}, O. Almaini$^{1}$, W. G. Hartley$^{1}$, R. J. McLure$^{2}$, J. S. Dunlop$^{2}$, S. Foucaud$^{3}$,
 \newauthor C. J. Conselice$^{1}$, C. Simpson$^{4}$, M. Cirasuolo$^{2}$ and E. J. Bradshaw$^{1}$\\
$^{1}$University of Nottingham, School of Physics \& Astronomy, Nottingham, NG7 2RD \\
$^{2}$Institute for Astronomy, University of Edinburgh, Royal Observatory, Edinburgh EH9 3HJ \\
$^{3}$Department of Earth Sciences, National Taiwan Normal University, No. 88, Section 4, Tingzhou Road, Wenshan District, Taipei 11677, Taiwan \\ 
$^{4}$Astrophysics Research Institute, Liverpool John Moores University, Twelve Quays House, Egerton Wharf, Birkenhead CH41 1LD  \\
}

\begin{document}

\date{Accepted 2010 December 20. Received 2010 December 8; in original form 2010 August 4}

\pagerange{\pageref{firstpage}--\pageref{lastpage}} \pubyear{2011}

\maketitle

\label{firstpage}

\begin{abstract}
We present a study of galaxy environments to z$\sim$2, based on a
sample of over 33,000 K-band selected galaxies detected in the UKIDSS Ultra Deep
Survey (UDS).  The combination of infrared depth and area in the UDS
allows us to extend previous studies of galaxy environment to $z>1$
without the strong biases associated with optical galaxy selection.
We study the environments of galaxies divided by rest frame $(U-B)$
colours, in addition to `passive' and `star-forming' subsets based on
template fitting.  We find that galaxy colour is strongly correlated
with galaxy overdensity on small scales ($<1$~Mpc diameter), with red/passive
galaxies residing in significantly denser environments than
blue/star-forming galaxies to $z\sim 1.5$. On smaller scales
($<0.5$~Mpc diameter) we also find a relationship between galaxy luminosity and environment, with the most luminous blue
galaxies at $z\sim1$ inhabiting environments comparable to red, passive
systems at the same redshift.  Monte Carlo simulations demonstrate
that these conclusions are robust to the uncertainties introduced by
photometric redshift errors.

\end{abstract}


\section{Introduction}

It has long been known that the properties of galaxies depend on the
environment in which they are located. Elliptical,
non-star-forming galaxies occupy more dense regions of space than
star-forming, disc-dominated galaxies, giving rise to the so-called
morphology-density relation (\citealt{Oemler};
\citealt{Dressler}). The physical origin of this relation is still
subject to debate, with disagreement mainly centering on whether the
relation arises due to internal or external processes (nature vs
nurture). 

Most recent low redshift studies (\citealt{Kauffmann};
\citealt{Balogh}) utilise the Sloan Digital Sky Survey (SDSS) or the
Two-degree-Field Galaxy Redshift Survey (2dFGRS) to conduct
statistical investigations of galaxy environments. \citet{Kauffmann}
constrained the specific star formation rate (SSFR) using the 4000\AA
~break and found that the SSFR (and nuclear activity) depend most
strongly on local density, from  star-forming galaxies at low
densities to predominantly inactive systems at high densities. \par Studies by
\citet{van Der Wel} and \citet{Bamford} found that structure, colour
and morphology are mainly dependent on galaxy mass but that at fixed
mass, colour and, to a lesser extent, morphology are sensitive to
environment. Studies of H$\alpha$ \citep{Balogh} found that it's
strength does not depend on environment but that the fraction of
galaxies with equivalent width, $W_0(H\alpha)>$4\AA ~is
environmentally dependant, decreasing with increasing density. They
also noted that emission line fraction appears to depend on both the
local environment ($\sim$ 1Mpc) and on the large scale structure
($\sim$ 5Mpc).

Studies at higher redshifts (z $\sim$ 1) have used surveys such as
DEEP2 \citep{Davis} and VVDS \citep{LeFevre}. DEEP2 investigations
(\citealt{Cooper06}; \citealt{Cooper07}) used the projected third
nearest neighbour statistic, studying galaxy properties and the
colour-density relation respectively. They concluded that there is a
strong dependence on rest frame $(U-B)$ colour, with blue galaxies
occupying lower density regions but showing a strong increase in mean
local density with luminosity at $z\sim 1$. This they conclude is consistent
with the rapid quenching of star formation by AGN or supernova feedback, as
ram pressure stripping, harassment and tidal interactions, which occur
preferentially in clusters, would be insufficient to explain these
findings. \cite{Cooper07}  also observed that the fraction of
galaxies on the red sequence increases with local density, 
as in the
local Universe, but  this weakens with redshift and disappears
by  $z\sim1.3$. 

The VIMOS VLT Deep Survey (VVDS ) investigated the redshift and
luminosity evolution of the galaxy colour-density relation up to
z$\sim$1.5 \citep{Cucciati}. In agreement with \cite{Cooper07} they
found that the local colour-density relation progressively weakens and
possibly reverses in the highest redshift bin (1.2$<$z$<$1.5). This
may imply that quenching of star formation was more efficient in high
density regions. The VVDS team also observed that the colour-density
relations depend on luminosity and found that at fixed luminosity
there is a decrease in the number of red objects as a function of
redshift in high density regions. This implied that star formation
ends at earlier cosmic epochs for more luminous/massive galaxies,
which is consistent with downsizing \citep{Cowie}. We note, however,
that the VVDS survey is based on optical $I$-band selection, and as
such will be strongly biased against red, passive galaxies at
$z>1$. Conclusions from deep K-selected samples suggest that the galaxy colour bimodality is present to at least z$\sim$1.5 (e.g. \citealt{Cirasuolo}) and may be still be present at z$\sim$2 (\citealt{Cassata08}; \citealt{Kriek08}; \citealt{Williams09}). Furthermore red galaxies have been seen to strongly cluster at z$>$1.5 (\citealt{Daddi03}; \citealt{Quadri07}; \citealt{Hartley08}; \citealt{Hartley}) which suggests that a colour-density relation may also exist at these higher redshifts.

A number of physical processes may be responsible for the observed
environmental trends. Mergers or tidal interactions can tear galactic
discs apart and are likely to play an important role in forming the
most massive galaxies observed today (\citealt{Toomre};
\citealt{Farouki}).  Other processes such as gas stripping can
severely reduce the star formation rate by removing the cold gas from
galaxies falling into massive dark-matter halos (\citealt{Gunn}; \citealt{Dekel}). Feedback
is also thought to play a major role, either from AGN or supernovae
(e.g. \citealt{Benson05}; \citealt{Springel}). These processes may heat or eject
the gas within galaxies and thus effectively terminate any further
star formation, which can rapidly lead to the build-up of the galaxy
red sequence. Finally infalling cold gas in low mass dark matter halos
may fall directly onto the galaxy, whereas in high mass halos the gas
is thought to be heated by shocks and therefore remains supported
(\citealt{White}; \citealt{Birnboim}). A key goal of observational
extragalactic astronomy is to disentangle which of these processes are
responsible for establishing the bimodal galaxy populations observed
in the local Universe.

With the recent advent of deep, wide-field infrared imaging in the
UKIDSS UDS we can now extend studies of galaxy environments to
z$>$1. Selection in the infrared avoids the major biases against dusty
and/or evolved stellar populations, allowing us to investigate whether
correlations observed at low redshift also occur at high redshift and
how these change over time. The large contiguous area of this survey
also allows us to probe a wide range of environments using large
samples of galaxies.

The paper is structured as follows:
\S 2 outlines the data and selection criteria  used in this work. \S 3
discusses the method  used to estimate galaxy environments. The results are
then presented in \S 4 and \S 5, with \S 6
summarising our conclusions. Throughout this paper we assume a
$\Lambda$CDM cosmology with $\Omega_{m}$=0.3, $ \Omega_{\Lambda}$=0.7
and $H_{0}$=71 km s$^{-1}$ Mpc$^{-1}$.

\section[]{Data and sample selection}

\subsection{The UKIDSS Ultra Deep Survey}

\hspace{5mm} This work has been performed using the third data release
(DR3) of the UKIRT (United Kingdom Infra-Red Telescope) Infrared Deep
Sky Survey, Ultra-Deep Survey (UKIDSS, UDS; \citealt{Lawrence};
Almaini et al in prep). The UKIDSS project consists of 5 sub-surveys
of which the UDS is the deepest, with a target depth of K=25 (AB) 
over a single 4-pointing mosaic of the Wide-field camera (WFCAM,
\citealt{Casali}), giving the UDS an area of 0.88 x 0.88 degrees. The
5$\sigma$, AB depths within 2$''$ apertures for the J, H and K-bands
are 23.7, 23.5 and 23.7 respectively for the DR3, making it the deepest
near infrared survey over such a large area  at the time of
release.  For details of the stacking procedure, mosaicing, catalogue
extraction and depth estimation we refer the reader to Almaini et
al. (in prep.) and \citet{Foucaud}. The field is also covered by deep
optical data in the B, V, R, i$^{\prime}$ and z$^{\prime}$ -bands with
depths of $B_{AB}$=28.4, $V_{AB}$=27.8, $R_{AB}$=27.7, $i'_{AB}$=27.7
and $z'_{AB}$=26.7 from the Subaru-XMM Deep Survey (SXDS)
\citep[$3\sigma$, $2\arcsec$ diameter]{Furusawa}. Data from the {\it Spitzer}
Legacy Program (SpUDS, PI:Dunlop) reaching 5$\sigma$ depths of 24.2
and 24.0 (AB) at 3.6$\mu$m and 4.5$\mu$m respectively and U-band data
from CFHT Megacam ($U_{AB}$=25.5; Foucaud et al. in prep) are also utilised, which results in a co-incident
area  of 0.63 deg$^2$ after masking.

\begin{figure}
\begin{center}
\includegraphics[angle=0,width=250pt]{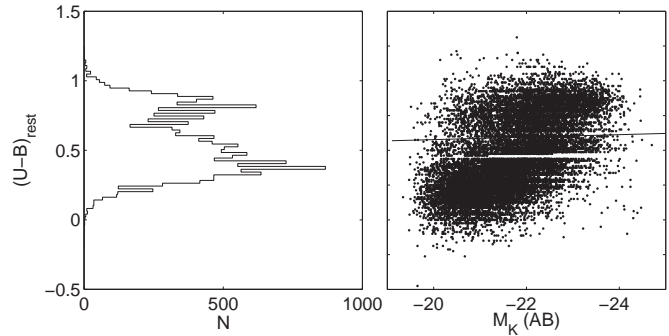}
\caption[$(U-B)$ vs absolute K-band magnitude between $0.75<z<1.75$]{Rest Frame $(U-B)$ colours for UDS galaxies between $0.75<z<1.75$. Shown as a histogram and as a function of
  absolute K-band magnitude. The division
  between red and blue is shown, using the red-sequence slope derived by
  Hartley et al. (2010).}
\label{UBMK}
\end{center}
\end{figure}

The galaxy sample used here is primarily based on selection in the
K-band image, upon which we impose a cut at $K_{AB}$=23 to
minimise photometric errors and spurious sources and also to ensure a
high level of completeness ($\simeq$100 per cent) and reliable
photometric redshifts (see \citealt{Cirasuolo10};
\citealt{Hartley}). Bright stars are trivially removed from the
combined catalogue by excluding objects on a stellar locus defined by
2-arcsec and 3-arcsec apertures, which is effective to  $K<18.1$.
The fainter stars are removed using a
$(B-z^{\prime})-(z^{\prime}-K)$ diagram \citep{Daddi} and the
criterion $(z^{\prime}-K)<0.3~(B-z^{\prime})-0.5$. These cuts and
careful masking of bright saturated stars and surrounding contaminated
regions leave 33,765 galaxies in our sample.

\subsection{Photometric Redshifts}

The magnitudes from the DR3 catalogue were used to determine
photometric redshifts and stellar ages by $\chi^2$ minimisation using
a large set of templates. This was performed with a code based largely
on the HYPERZ package \citep{Bolzonella} using both average local
galaxy SEDs and templates from the K20 survey. Six SEDs of observed
starbursts from \cite{Kinney} were also used to improve the
characterisation of young blue galaxies. This yeilded photometric redshifts 
with a $\delta$z/(1+z)=0.008 with a standard deviation of $\sigma$=0.034 after the exclusion of outliers (for more detail see
\citealt{Cirasuolo}; \citealt{Cirasuolo10} and references
within). This also provided rest frame magnitudes and colours, of which
the rest frame $(U-B)$ and $K$-band absolute magnitude are utilised
here. Sources with an unacceptable fit ($\chi^2>15$) are removed from
our sample as these are likely to be unreliable. This removes 4$\%$ of
the galaxy sample, the majority of these are either QSOs
(36\%), cross-talk (26\%) or the minor members of pairs or mergers
(23\%),  with the remainder consisting largely of objects with very low
surface brightness. The fraction of otherwise useful objects rejected
is therefore $<$0.6$\%$.

\subsection{Passive Sample}

To define a passive galaxy subset with minimal contamination from dusty star
forming objects we use a subset of galaxies outlined in 
\cite{Hartley}. Templates were used to fit either an instantaneous
burst parameterised by an age, or an exponentially decaying
star-formation rate parameterised by an age and $\tau$, the e-folding
time in the exponentially declining star-formation rate, such
that,
\begin{equation}
SFR = SFR_{0} \times  e^{-age/\tau} 
\end{equation} 
where $SFR$ is the star-formation rate at the time of observation and
$SFR_{0}$ was the initial value. We define a conservative passive sample as galaxies that are
simultaneously old (age$>$1Gyr) and have ongoing star formation with
$SFR\le 0.1 \% $ of $SFR_0$, and a star forming
sample with $SFR\geq10\% $ of $SFR_0$, with 3947 and 22,158 galaxies in
each sample respectively.\par
To define the red sequence we performed a $\chi^2$ minimisation to fit an equation of the form $(U-B)
= a \times  M_K + b$~ to the old, burst galaxies, defining the red sample to
be all galaxies within 3$\sigma$ of this fit (see \cite{Hartley}
for a more detailed description). In this work, to separate red and
blue galaxies, we use the red-sequence slope from \citet{Hartley} but
choose the division between the two populations to fit the minimum in
the overall colour bimodality (as shown in the histogram in Figure
\ref{UBMK}). This leads to a dividing line in the colour-magnitude
diagram as follows:
\begin{equation}
 (U-B)=-7.09 \times 10^{-3} M_K + 0.52
\end{equation}
This boundary was found to separate the red and blue populations effectively to z$\sim$1.75. At higher redshift the bimodality in galaxy colours is less clear, which may in part be due to photometric errors. A full examination of this issue and the evolution of the red sequence will be presented in Cirasuolo et al. (in prep). Previous studies have found evidence for an evolution in the location of the red sequence with redshift (e.g \citealt{Brammer}). For simplicity we choose not to model the red sequence in such detail and instead use the fixed colour selection boundary given above. We note, however, that using an evolving boundary made no significant difference to any of the conclusions presented in this work. 

Table \ref{NGal} shows the resulting number of red and blue galaxies assigned 
to each photometric redshift bin, including the conservative subsamples of 
passive and actively star-forming galaxies.

\section[]{Environmental Measurement}

We used two methods to calculate galaxy environment: Counts in an
Aperture and $n$th Nearest Neighbour. In both methods all the galaxies
within a photometric redshift bin are collapsed down onto a 2D plane and the
redshift information within the bin is not utilised any further. In
the aperture method  apertures of 1Mpc, 500kpc and
250kpc diameter (physical) are placed on each galaxy and the number of
galaxies within that aperture are counted (N$^{Aper}_{g}$). A
sample of $\sim$100,000 random galaxies are then put down in the
unmasked regions and the number of randoms within the aperture are
counted (N$^{Aper}_{r}$). The number of galaxies within the aperture
is then normalised to give the final density measurement,
$\rho/\rho_r$:
\begin{equation}
 \frac{\rho}{\rho_r}=\frac{N^{Aper}_{g}}{N^{Aper}_{r}} \times \frac{N^{Tot}_{r}}{N^{Tot}_{g}}
\end{equation}
where N$^{Tot}_{r}$ and N$^{Tot}_{g}$ are then total number of random
points and galaxies respectively, so that $\rho/\rho_r$=1 corresponds
to a density consistent with that of a random distribution of
galaxies. This method was chosen to be the basis of this work as it is
conceptually simple, and as concluded by \citet{Cooper05}, this
technique has a distinct advantage in fields masked by a large number
of holes. The nearest neighbour method would require the exclusion of
a large fraction of data close to holes and field edges.

The $n$th nearest neighbour method was first employed by
\citet{Dressler}, this calculates the distance to the $n$th nearest
galaxy, $D_{n}$ in Mpc and is expressed here as a surface density,
\begin{equation}
\Sigma_n = \frac{n}{\pi D_n^2} 
\end{equation}  
The surface density, $\Sigma_n$ is then renormalised such that,
\begin{equation}
\delta_n = {\frac{\Sigma_n}{\bar{\Sigma}}} 
\end{equation}
where $\bar{\Sigma}$ is the median density of galaxies within the field. To reduce the effect of the edges, the distance to the nearest edge was calculated and if this was less than the distance to the third nearest neighbour then the object was removed from the sample. This method was only used in this work to test the primary findings of the aperture method. The results are presented in the appendix.    

\section[]{Results}

Below we explore the relationship between galaxy colours and environment as a function of redshift. Relatively broad redshift bins are used to minimise the contamination due to photometric redshift errors. These sources of uncertainty are explored further in section 5. 

\begin{figure}
\begin{center}
\includegraphics[angle=0, width=250pt]{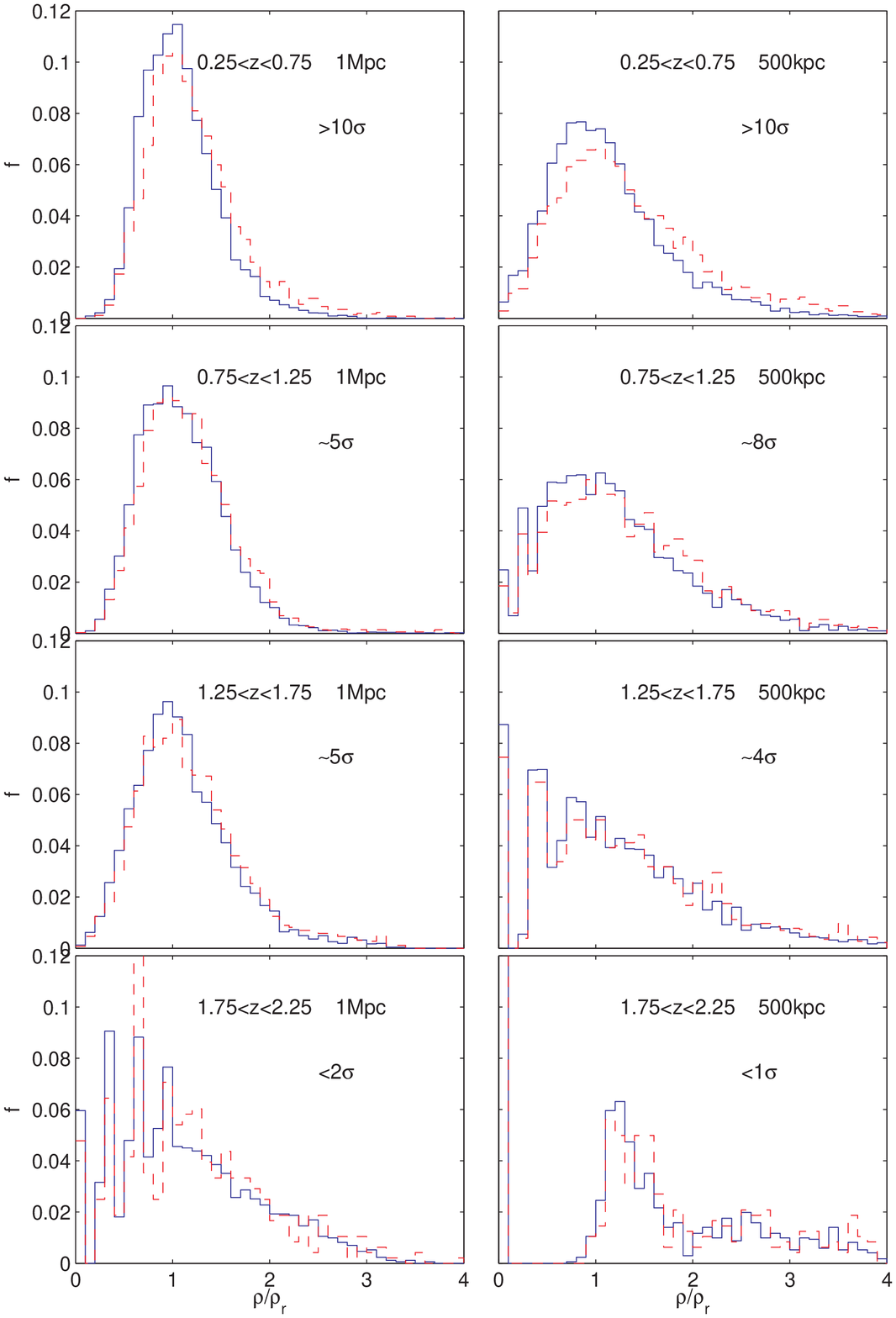}
\caption{Histograms of the density of galaxies within 1Mpc and 500kpc
  diameter apertures compared to a random sample for red (dashed line)
  and blue (thick line) galaxies. The $\sigma$ values
  are obtained by performing a KS test, representing the significance in rejecting
  the null-hypothesis that the samples are drawn from the same
  underlying population.}
\label{RBHist}
\end{center}
\end{figure}

\begin{figure*}
\begin{minipage}{150mm}
\begin{center}
\includegraphics[angle=0, width=400pt]{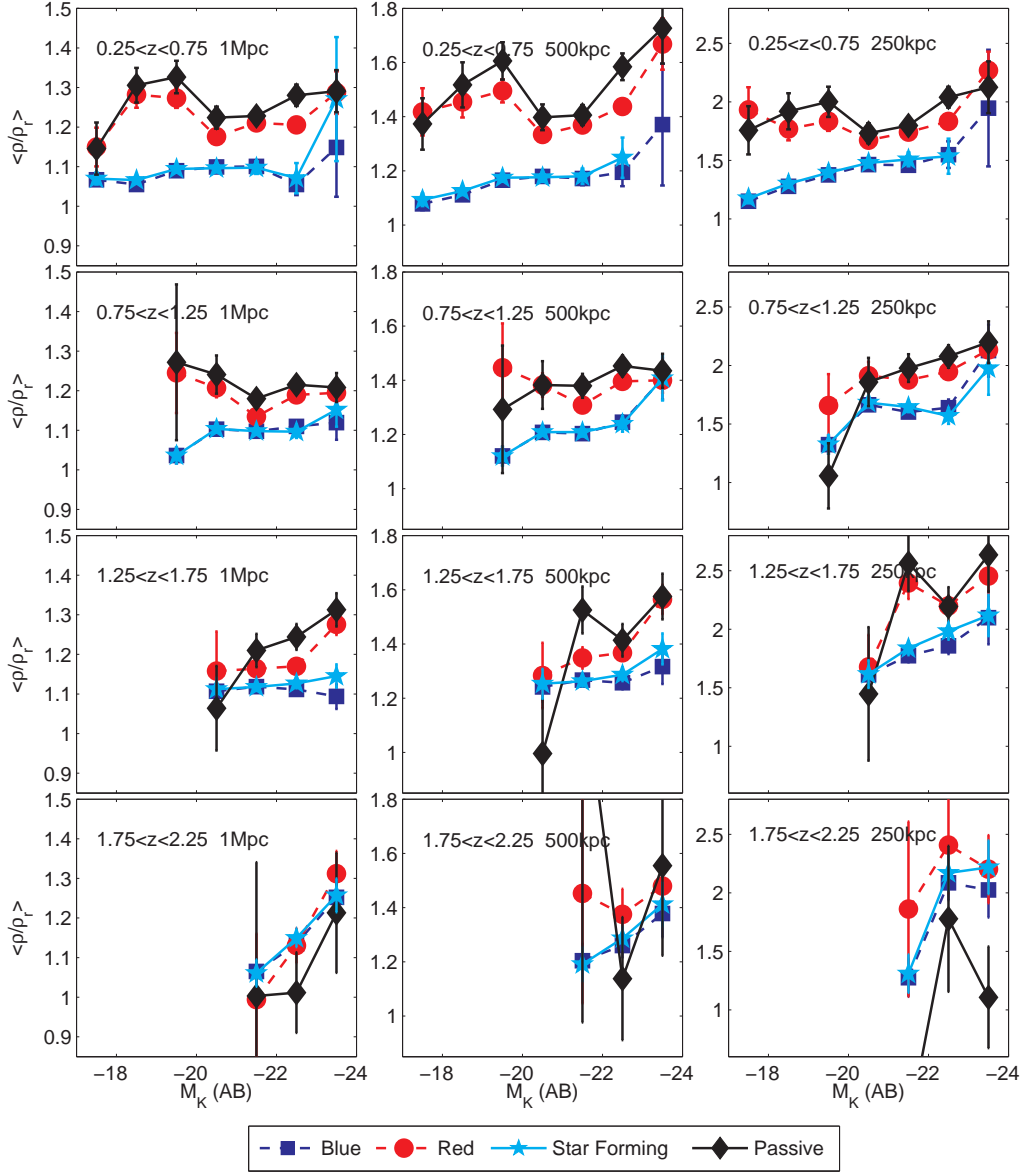}
\caption{The average galaxy overdensity as a function of K-band
  luminosity, displayed  in four redshift
  bins (top to bottom) and using projected apertures of diameter 1Mpc,
  500kpc and 250kpc (three columns). Galaxies are displayed in red,
  blue, passive and star-forming subsets, as defined in Section
  2. Note the change in scale for each column. The environments are defined 
so that $\rho/\rho_r$=1 corresponds to a density consistent with that of 
a random distribution of galaxies.}
\label{RBPass}
\end{center}
\end{minipage}
\end{figure*}

\begin{table*}
\begin{minipage}{150mm}
\begin{center}
\begin{tabular}{|c|c|c|c|c|}
\hline
 &   $0.25<z<0.75$ &  $0.75<z<1.25$ &  $1.25<z<1.75$ & $1.75<z<2.25$\\ \hline
Red & 3,469  & 3,500 & 2,565 & 481 \\ \hline
Blue & 7,367 & 7,255 & 4,674 & 1,711 \\ \hline
Passive & 1,623 & 1,395 & 722 & 87 \\ \hline
Star-forming & 6,938 & 6,741 & 5,048 & 1,817 \\ \hline
\end{tabular}
\end{center}
\caption{Table showing the number of galaxies in each sample and in each redshift range, including our conservative subsets of passive and actively star-forming galaxies.}
\label{NGal}
\end{minipage}
\end{table*}

\subsection{Red and Blue Environments}

Figure \ref{RBHist} shows histograms of galaxy environments within two
apertures of 1Mpc and 500kpc diameter. These are divided into four
redshift bins between $z=0.25$ and $z=2.25$ and into red and blue
populations. A KS-test was performed on the samples to assess the
difference between the red and blue populations, with all but the final
redshift bin showing highly significant differences between the two
populations. In the first three redshift bins we can exclude the null
hypothesis that the red and blue galaxies are drawn from the same
underlying population with a significance in excess of 5$\sigma$
significance, falling to $<$2$\sigma$ and $<$1$\sigma$ significance 
at $z>1.75$.

Figure \ref{RBPass} is a plot of the mean density of red, blue,
passive (black) and star forming (cyan) galaxies in bins of absolute
K-band magnitude. Error bars are derived from the error on the mean
density of galaxies within a given bin. Sources of error are explored
further in Section 5. The passive and star forming galaxies are
defined in Section 2. As before they are plotted in four redshift bins
but with an additional 250kpc aperture. This plot illustrates that red
and/or passive galaxies reside in significantly denser environments than 
blue and/or star-forming galaxies from the present day to  $z\sim
1.5$, and this difference is apparent at all luminosities. This is
comparable to what has been found in the local universe by other
studies (\citealt{Kauffmann}; \citealt{van Der Wel}).

Figure \ref{RBPass} also shows that passive galaxies (shown in black)
follow a similar density profile to red galaxies but are on average in
slightly denser environments. This supports the conclusion that
passive galaxies within the red population are responsible for the
enhanced environments compared to blue star-forming objects. The 
environments of galaxies that were red but not in the strict `passive' sample 
were also investigated and these were found to lie in-between the red and blue 
galaxy environments, as would be expected (these are not shown for clarity). The
actively star-forming galaxies exhibit the same environmental
dependence as the blue galaxies, following the same luminosity-density
profile. \par

In addition to the clear separation of red and blue galaxies, we also
find a general trend of increasing galaxy density with luminosity,
particularly for blue galaxies and on smaller scales. Inspecting the
two intermediate redshift bins in Figure \ref{RBPass}, on scales below
$500$kpc we find that the most luminous blue galaxies appear to
inhabit environments approaching those of red/passive galaxies. These
results are consistent with the findings of \cite{Cooper07}, who
observed a strong increase in local density with luminosity for blue
galaxies at $z\sim 1$. Our results appear to extend these findings to
higher redshift, suggesting that the epoch $1<z<2$  represents a key
period of transformation of massive galaxies from the blue cloud onto
the passive red sequence.

\begin{figure}
\begin{center}
\includegraphics[angle=0,
  width=250pt]{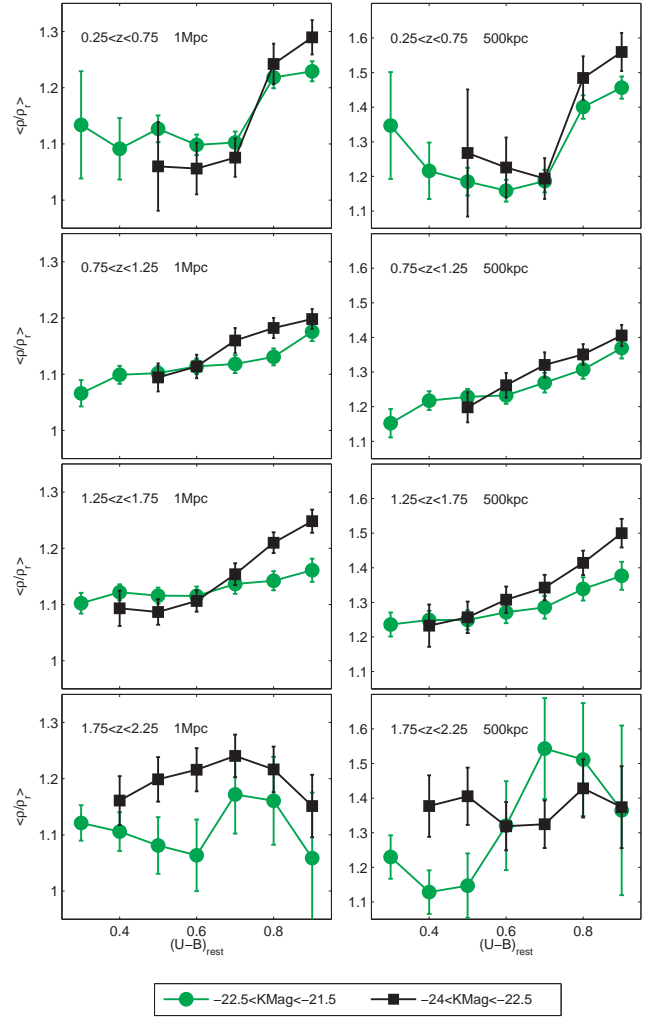}
\caption{The average galaxy density in a 1Mpc and 500kpc diameter apertures compared to a random sample as a function of $(U-B)$ rest frame colour. We separate high (black) and low (green) luminosity subsets. Note the change in y-axis values between the two different aperture sizes.}
\label{Colour}
\end{center}
\end{figure}

\subsection{Colour-Density Relation}

In Figure \ref{Colour} we display average density as a function of
$(U-B)$ rest-frame colour, with faint (-22.5$<M_K<$-21.5) and luminous (-24$<M_K<$-22.5) objects shown in
green and black respectively. This shows a clear trend for galaxies
below z$\sim$2, with redder galaxies occupying denser average
environments. In the lowest redshift bin this change occurs very abruptly at the boundary in the colour bimodality at
$(U-B)_{rest}\sim$0.7 (see Figure \ref{UBMK}). There is also a general trend for more luminous galaxies to occupy denser environments.

\begin{figure}
\begin{center}
\includegraphics[angle=0, width=250pt]{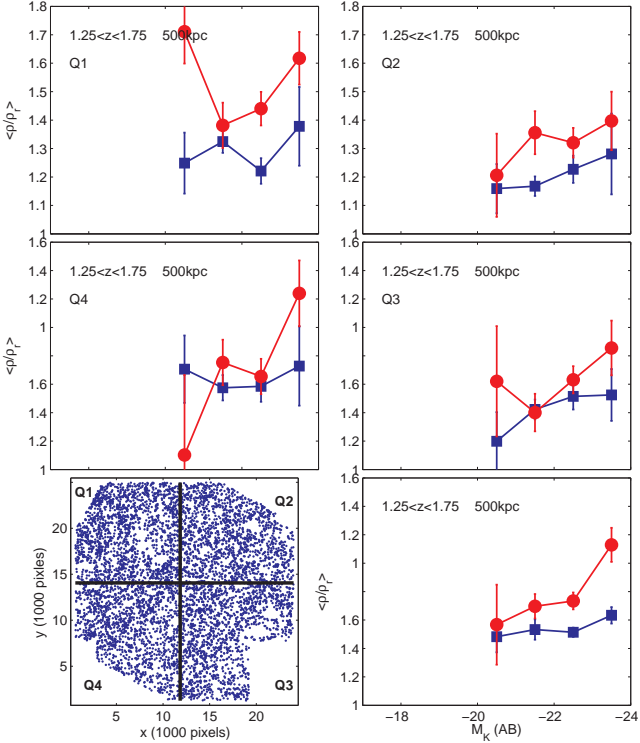}
\caption{Showing the density of red and blue galaxies against K-band
  luminosity in four quadrants of the UDS field shown in the bottom
  left, using median values of x and y to calculate the position of
  the centre. The overall result for the field at 1.25$<$z$<$1.75 is
  shown in the bottom right with the error bars calculated from the
  error on the mean of the four quadrants, to give an estimate of the
  cosmic variance.}
\label{CosVarz3}
\end{center}
\end{figure}

In summary, we have found strong evidence that red galaxies are on
average found in denser local environments than blue galaxies,
extending the comparison to a higher redshift than any previous study.
\citet{Cooper07} probed up to $z\sim1.3$, finding that red and blue
galaxies at this redshift occupy indistinguishable environments. In
contrast we find that there is a significant difference between red
and blue galaxies between $1.25<z<1.75$ (see Figures \ref{RBHist}
and \ref{RBPass}). This difference is likely to be due to our method
of selecting galaxies from deep infrared imaging. Very red, passive
galaxies will have been missed from the R-band 
selected candidates used in the DEEP2 study. This selection effect can be 
quantified using our UDS sample. Considering only the most luminous galaxies 
(M$_K$$\leq$-22.5) in the redshift range 1$<$z$<$1.3, if we impose the selection criteria 
of DEEP2 (R$_{AB}$$\leq$24.1) we find that 75\% of red galaxies would be missed compared to 
only 20\% of the blue sample. At fainter luminosities virtually no red/passive 
galaxies are selected.

\section{Investigating Sources of Error} 

In this section we conduct a number of tests to investigate various
source of systematic uncertainty. We attempt to quantify the
cosmic variance in our data by dividing the survey into four
quadrants. A volume limited study of the brighter galaxies is also
performed, to allow a clearer comparison between different redshift
bins. Finally we investigate the effects of photometric redshift
errors using Monte Carlo simulations.

\subsection{Cosmic Variance}
To attempt to quantify the effects of cosmic variance on our results
we divide the UDS field into four quadrants.  The median values of the
$x$ and $y$ positions of the galaxies were used to divide the field,
to ensure that the number of galaxies in each quadrant was comparable.
The results for the galaxies in the redshift range 1.25$<$z$<$1.75 are
shown in Figure \ref{CosVarz3}.  The average density versus K-band
absolute magnitude is shown for the four quadrants, with the bottom
left panel showing the projected distribution in the field. The lower
right panel shows the original result from Figure 3 (red and blue
only) but with the error bars calculated from the error on the mean of
the densities in the four quadrants in each magnitude bin. It can be
seen that the distinction between red and blue galaxy environments
remains, although there are clearly significant variations across the
field.  We conclude that while field variance is clearly present in
these data the primary findings  remain robust. Similar conclusions
were drawn from other redshift bins and on other scales, which are not
shown in the interests of brevity.

As a cautionary remark, however, we note that strictly speaking we have only
tested the {\em internal} field variance, since despite the relatively
wide field of the UDS ($50\times 50$ comoving Mpc at $z\sim 1$) we may
nevertheless be prone to large-scale cosmic variance due to unusual
superstructures (Somerville et al. 2004). Testing against such effects
will require further wide-area infrared surveys, such as the forthcoming
VIDEO and Ultra-VISTA surveys.

\begin{figure}
\begin{center}
\includegraphics[angle=0, width=250pt, trim = 0mm 0mm 0mm 0mm, clip]{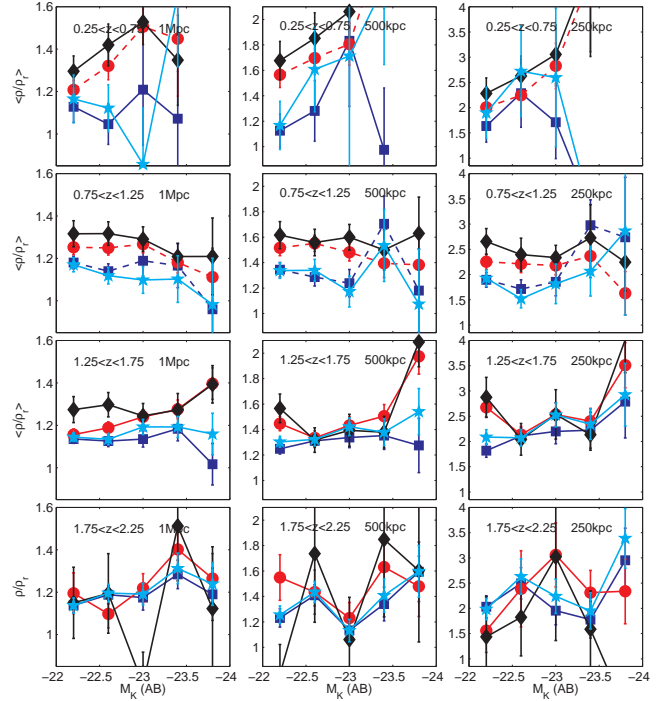}
\caption{As Figure 3, but using only the brightest galaxies with M$_K\leq$-22 (AB) as  tracers of galaxy density. Note the change in y-axis values between the three different aperture sizes}
\label{RB24}
\end{center}
\end{figure}

\subsection{Volume-Limited Sample}
Our measurement of environmental density is based on a flux limited
survey, so by definition we are using fainter galaxies to define
environments at low redshift.  To investigate the influence of this
systematic effect we reduced the sample to only the brightest galaxies
with $M_K\leq$-22 (AB) and repeated the environmental analysis using this
volume-limited sample.  This allows the galaxies to be used as tracers
in the measure of environment over the same range in luminosity at all
redshifts, to allow a fairer comparison between epochs. The results
are shown in Figure \ref{RB24}, where the average density versus
K-band luminosity is shown for comparison with the flux-limited study
in Figure \ref{RBPass}.  As expected, the volume-limited study is much
noisier (particularly at low redshift) but the results are consistent
with the primary findings of Section 4. Red and/or passive galaxies
are found to occupy denser environments on average to $z\sim 1.5$.

\begin{figure}
\begin{center}
\includegraphics[angle=0, width=250pt]{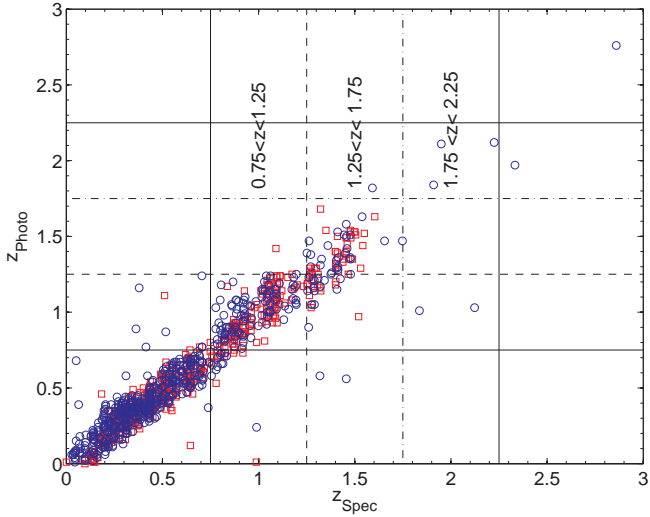}
\caption{Plot of the reliable spectroscopic redshifts against photometric redshifts, highlighting the three highest redshift bins, 0.75$<$z$<$1.25, 1.25$<$z$<$1.75 and 1.75$<$z$<$2.25, with AGN and radio galaxies removed. The red squares and blue circles indicate the red and blue galaxies respectively, as defined in section 2.2.}
\label{zvsz}
\end{center}
\end{figure}

\begin{figure}
\begin{center}
\includegraphics[angle=0, width=250pt]{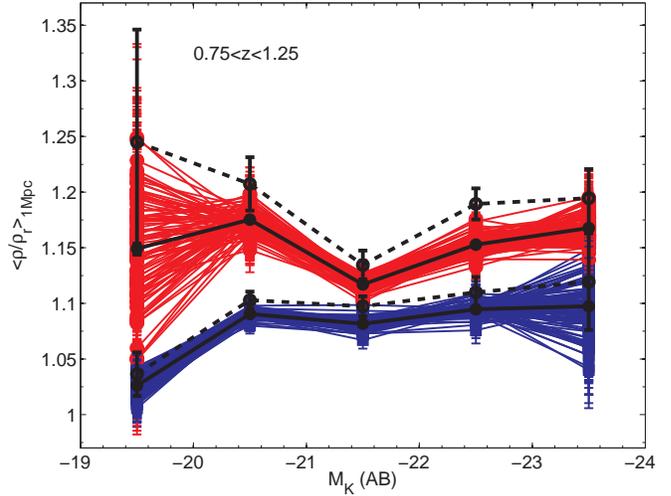}
\caption{Monte Carlo simulation showing the effect of the photo-z errors on our measurement of galaxy overdensity, based on 100 simulations. The thick black line shows the average of the simulations and the dashed black line shows the original result for 0.75$<$z$<$1.25 from Figure \ref{RBPass}.}
\label{MCz2n3}
\end{center}
\end{figure}

\begin{figure}
\begin{center}
\includegraphics[angle=0, width=250pt]{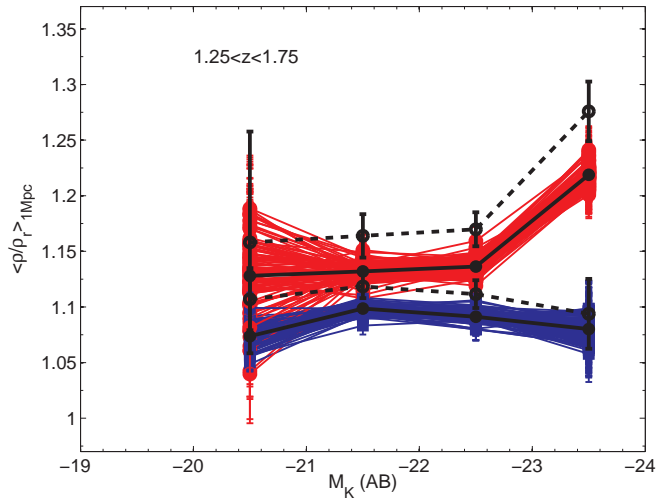}
\caption{As Figure \ref{MCz2n3} but for the 1.25$<$z$<$1.75 bin.}
\label{MCz4n5}
\end{center}
\end{figure}

\subsection{Monte Carlo Simulations}

Monte Carlo simulations were conducted to investigate the effect of 
photometric redshift errors on our results, explicitly allowing for the effects of outliers and catastrophic errors. Such errors would
generally dilute differences in galaxy environments, but could
potentially introduce fake overdensities if a large fraction of
low-redshift structure is incorrectly assigned to a higher redshift bin.
We note, of course, that the effects of photo-z errors are already
present in the data, so these simulations can only provide an indication
of the magnitude of the shifts due to these effects. 

The distribution of spectroscopic versus photometric redshifts used in
this field are shown in Figure \ref{zvsz}, with spectroscopic redshifts
derived from a variety of sources (\citealt{Yamada}; \citealt{Simpson};
Akiyama et al., in prep; Simpson et al., in prep; Smail et al., in
prep) as outlined in Cirasuolo et al. (2010). We use the distribution in each redshift bin (higher redshift bins indicated in Figure \ref{zvsz}) to estimate the fraction of galaxies incorrectly
reassigned from one redshift bin to another.  For simplicity two
categories were adopted: `catastrophic' errors were defined as those
with $\delta z>0.25$, while others are classified as `normal'.  In the
absence of catastrophic errors we expect $>99.9 \%$ of galaxies to
show $\delta z<0.25$ at $z\sim 1$ (Cirasuolo et al. 2010).  The
resulting fractions were then randomly applied to the full sample,
shifting galaxies into new bins to simulate the effect of photometric
redshift errors (`normal' and `catastrophic'). 

This was repeated 100 times per redshift bin. Results are displayed
for the redshift bins $0.75<z<1.25$ (Fig.\ref{MCz2n3}) and
$1.25<z<1.75$ (Fig.\ref{MCz4n5}). These simulations demonstrate that
photometric redshift errors tend to dilute the distinction between the
red and blue populations rather than introduce fake differences.  The
resulting additional sources of error in galaxy density are comparable
to the errors on the mean displayed previously. Only in highest
redshift bin ($z>1.75$) do the simulated populations introduce
significant extra scatter (not shown). Given these findings, and the
lack of spectroscopic redshifts at $z\sim 2$, we urge caution in
interpreting our preliminary results in this highest redshift bin.  At
lower redshifts, however, the environmental differences appear robust.

\begin{figure} \begin{center} \includegraphics[angle=0,
width=250pt]{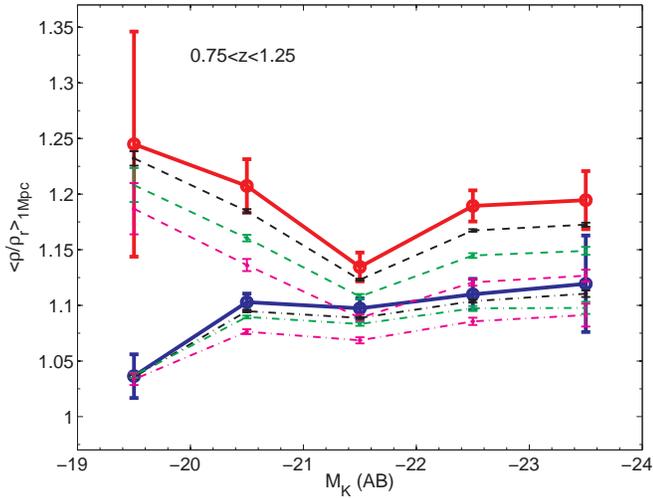}
\caption{The results of Monte Carlo simulations based on the 
extreme scenario in which all incorrectly assigned photometric
redshifts lead to blue galaxies being incorrectly classified as red,
and vice versa. The original measures of environment for red galaxies and blue
galaxies as a function of luminosity are shown as thick red and blue
lines. Black dashed (dot-dashed) lines show the results of Monte Carlo
simulations for the red (blue) populations respectively. The corresponding
green and magenta lines illustrate the influence of doubling and tripling
the number of incorrectly assigned galaxies.}
\label{MCColours}
\end{center}
\end{figure}

Noting that we have only 97 spectroscopic redshifts in the redshift bin 1.25$<z<$1.75 (with 64 between 1.25$<z_{photo}<$1.75), where environmental differences appear to be very significant, we ran additional simulations using only half of the spectroscopic sample. The results were very similar to Figure \ref{MCz4n5}, indicating that we are not suffering from low number statistics.

As an additional test, we conduct further Monte Carlo simulations to
investigate the effects of contamination of the blue galaxies by the red
population and vice versa. These simulations were motivated by the
worry that incorrect redshifts may inevitably lead to errors in the
resulting restframe $(U-B)$ colours. We adopt an extreme approach by
simulating 100\% misclassification, using the contaminating fractions
outlined above but moving only blue galaxies into the red sample and
vice versa.  We find that these simulations have the effect of
diluting the environment for both red and blue simulations, but once
again are insufficient to influence any of the results previously
presented. To investigate further, the contaminating fractions are
artificially increased to double and triple the maximal contamination
indicated by spectroscopic redshifts. These only serve to further
reduce  the average galaxy densities.  Results for the $0.75<z<1.25$ bin
are illustrated in Figure \ref{MCColours}, and similar results are
found for all other redshift bins.\par
 We conclude that photometric redshift
errors (including catastrophic errors) have only served to dilute the differences in environment, but
an unrealistic proportion of misclassified galaxies are required to
significantly influence our primary findings.

\section{Conclusions}

In this paper we have presented a study of the environment of over
33,000 K-band selected galaxies between 0.25$<$z$<$2.25. We have used
a simple measurement of projected galaxy overdensities to study the environments of
galaxies divided into red and blue based on rest frame $(U-B)$ colours, as
well as using `passive' and `star-forming' galaxies from template
fitting. We attempt to quantify the effects of cosmic variance,
photometric redshift errors and flux-limit  biases on the resulting
environmental measurements. We find the following principle results:

\begin{enumerate}
\item We find a strong relationship between rest-frame $(U-B)$ colour
  and galaxy environment to $z\sim 1.5$, with red galaxies residing in
  significantly denser environments than blue galaxies on scales below
  $1$Mpc.  These results are robust to the effects of field variance,
  flux-limit biases and photometric redshift errors. The environments
  appear indistinguishable by $z\sim 2$, but the current lack of
  spectroscopic redshifts at this epoch does not allow a robust test
  of this tantalising signal at present.
\item Selecting `passive' and `star-forming' galaxies using template
  fitting yields consistent results, with the passive subset occupying
  slightly denser environments than the global red population at all
  epochs.  We conclude  that the passive subset of the red
  galaxies are responsible for the enhanced environments compared to blue galaxies.
\item On small scales ($<0.5$~Mpc) we find evidence for a positive
  correlation between galaxy K-band luminosity (a good proxy for
  stellar mass) and local density.  This trend is particularly clear
  for star-forming and blue galaxies, with the most luminous blue
  galaxies at  $z\sim 1-2$ showing average environments
  comparable to passive red galaxies.  These findings appear in very
  good agreement with the findings of \citet{Cooper07} at $z\sim 1$.
  These results are consistent with the identification of high-z
  galaxies in transition to the red sequence in the densest
  environments.
\end{enumerate}\par

To improve on this study, we note that the UDS imaging
programme  will continue at UKIRT until 2012 and ultimately probe $\sim1$
magnitude deeper than the data used in this work.  This will be
sufficient to detect $L^*$ galaxies to $z\sim 5$ in addition to tens
of thousands of sub-$L^*$ galaxies at lower redshift. Furthermore, an
ongoing ESO Large Programme (UDSz) will soon transform the UDS project
with the addition of $\sim 3000$ galaxy spectra. This will increase
the number of UDS spectroscopic redshifts at $z>1.5$ 
by an order of magnitude, which we expect to dramatically improve the
reliability of photometric redshifts and allow us to extend our study
of galaxy environments and large-scale structure to the crucial epoch
when the galaxy red sequence is first established.

\section*{Acknowledgements}

We are grateful to the staff at UKIRT for operating the telescope with 
such dedication. We also thank the teams at CASU and WFAU for processing 
and archiving the data. RWC would like to thank the STFC for their financial support. JSD acknowledges the support of the Royal Society via a Wolfson Research Merit award, and also the support of the European Research Council via the award of an Advanced Grant.

\section*{Appendix: Nearest Neighbour method}

\begin{figure}
\begin{center}
\includegraphics[angle=0, width=250pt]{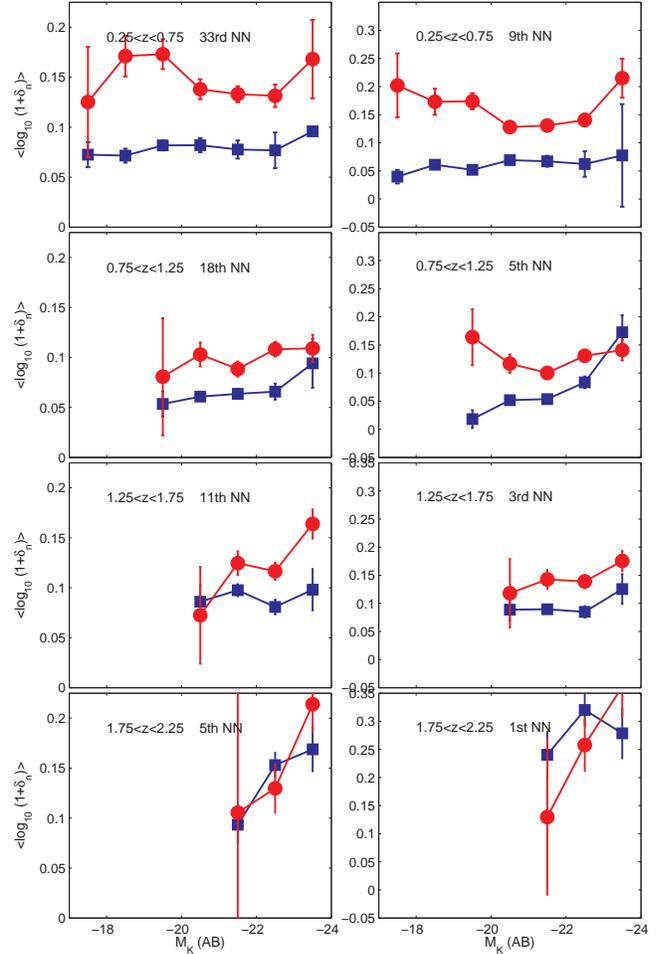}
\caption{The nearest neighbour density of red and blue galaxies with
  the nearest neighbour chosen so that the mean distance probed was a diameter of 1Mpc
  and 500kpc in each redshift bin.}
\label{ApNN}
\end{center}
\end{figure}

In this appendix we use the nth nearest neighbour method as an
alternative environmental estimator, to provide a test of the results
presented earlier and to allow comparison with previous work
(e.g. \citealt{Cooper07}). The nth nearest neighbour was calculated
for all the galaxies as a function of absolute magnitude, using the
method outlined in Section 3. The value of n is chosen to probe the
same mean scale as the apertures method.  The results are shown in Figure
\ref{ApNN}), which appear globally consistent with those derived by
the aperture method.

\bsp

\label{lastpage}

\end{document}